\documentclass{Interspeech2024}
\usepackage{url}
\usepackage{authblk}
\interspeechcameraready

\title{MUSE: Flexible Voiceprint Receptive Fields and Multi-Path Fusion Enhanced Taylor Transformer for U-Net-based Speech Enhancement}

\name[affiliation={1,*}]{Zizhen}{Lin}
\name[affiliation={2,*}]{Xiaoting}{Chen}
\name[affiliation={1,*}]{Junyu}{Wang}

\address{
  $^1$School of Electronic Information, Sichuan University, China\\
  $^2$School of Software, Yunnan University, China \\
  \small{$^*$Authors contributed equally to this work.}
}
\email{linzizhen17@163.com, 12022219155@mail.ynu.edu.cn, junyu\_wang@stu.scu.edu.cn}

\keywords{speech enhancement, multi-path enhanced taylor transformer, simplified channel attention, deformable embedding, u-net}

\begin{document}

\maketitle

\begin{abstract}
 Achieving a balance between lightweight design and high performance remains a challenging task for speech enhancement. In this paper, we introduce Multi-path Enhanced Taylor (MET) Transformer based U-Net for Speech Enhancement (MUSE), a lightweight speech enhancement network built upon the U-Net architecture. Our approach incorporates a novel Multi-path Enhanced Taylor (MET) Transformer block, which integrates Deformable Embedding (DE) to enable flexible receptive fields for voiceprints. The MET Transformer is uniquely designed to fuse Channel and Spatial Attention (CSA) branches, facilitating channel information exchange and addressing spatial attention deficits within the Taylor-Transformer framework. Through extensive experiments conducted on the VoiceBank+DEMAND dataset, we demonstrate that MUSE achieves competitive performance while significantly reducing both training and deployment costs, boasting a mere 0.51M parameters.
\end{abstract}

\section{Introduction}

Speech enhancement algorithms, also known as speech denoising algorithms, constitute a pivotal task within the realm of speech processing. Their applications extend to improving the quality of recorded audio, enhancing call quality, and augmenting the accuracy of speech recognition systems. In recent years, the flourishing landscape of deep learning has given rise to numerous advanced algorithms in the field of speech enhancement \cite{bulut2020low,tran2020single,wisdom2019differentiable,twosunet}. Noteworthy is the fact that these deep learning algorithms exhibit a remarkable capability to suppress complex and non-stationary noise.

Recently, researchers have explored diverse approaches in the realm of speech enhancement, various domains such as the time domain \cite{pascual2017segan,park2022manner,wang2021tstnn,defossez2019demucs}, time-frequency domain \cite{wang23DPCFCS_interspeech,DB-AIATinterspeech12,mpsenet_interspeech,cmgan_interspeech}. T-F methods apply short-time Fourier transform (STFT), which transform speech signal into time-frequency spectrum. In pursuit of a more comprehensive extraction of magnitude and phase features from speech signals, a variety of multi-domain fusion methods have been applied to Speech Enhancement (SE), by extract feature of magnitude and phase separately \cite{mpsenet_interspeech}. 

For previous methods, a higher number of channels (64) is often required to achieve peak performance. The popular Two-Stage (TS) \cite{wang2021tstnn} structural network in speech enhancement demonstrates strong competitiveness. The TS structure typically performs downsampling only once along the frequency dimension due to the large size of the feature maps. This implies that such structures require greater memory and computational resources, particularly as the length of the speech increases, necessitating significant GPU memory during both training and inference stages. Training or deploying such speech enhancement models on devices with limited GPU memory presents evident challenges. To mitigate GPU memory consumption for the same length of speech, the simplest approach is to reduce the number of channels to decrease model size. However, for speech enhancement tasks, the model needs to map speech features to a high-dimensional space; insufficient channel numbers often result in decreased ability to discriminate noise features from speech features, potentially leading to significant performance degradation. We retrained the state-of-the-art (SOTA) model MP-SEnet \cite{mpsenet_interspeech} with 16 channels as a baseline, subsequently referred to as MP-SEnet-16. Experimental results indicate that while MP-SEnet-16 significantly reduces GPU memory requirements and parameter count compared to its 64 channels versions, the model's performance also significantly deteriorates. Therefore, for scenarios with low channel counts, we consider continuing to use the TS structure as suboptimal.
Further more, for voiceprint information in speech, the feature shape often resembles an elongated crescent, contrasting with normal conv kernels that are fixed in a square shape. Hence, learning features from those specific shaped voiceprint becomes challenging and inefficient. 

We believe that the reasons limiting the performance boundaries of small-sized speech enhancement systems can be summarized as follows:

\begin{itemize}
\item We posit that the TS-Conformer structure is suboptimal for light weight speech enhancement networks, as the serial architecture fails to establish a robust information flow and hinders the effective integration of low-level and high-level features across different layers when network is shallow. Consequently, this limitation results in inadequate representational capacity of the network. Furthermore, the elevated computational costs impose constraints on the quantity of conformer blocks that can be feasibly employed.
\item For idiosyncratic voiceprints, standard convolutions exhibit limited adaptability in their receptive fields, rendering them inefficient in capturing distinctive features inherent in the vocal characteristics.
\item While vanilla Transformer-based approaches, such as multi-head self-attention (MSA)  \cite{msa-se} have demonstrated improved performance and the ability to capture intricate features, a critical consideration arises regarding the trade off between performance and computational cost. Employing a more efficient MSA mechanism and superior model architecture can be considered.
\end{itemize}
In this study, we introduce a novel U-Net based model MUSE for light weight speech enhancement. To ensure a resilient flow of information and impede the efficient integration of features spanning both low-level and high-level across various layers, the TS-Conformer \cite{mpsenet_interspeech} was omitted in favor of a U-Net \cite{u-net} architecture that incorporates deformable convolutions.
To effectively and flexibly learn voiceprint feature and high-frequency information, concomitantly facilitating inter-channel information exchange, we propose a novel Multi path Enhanced Taylor (MET) attention mechanism. We substitute Taylor-Multi-head Self Attention (T-MSA) for MSA, significantly reducing computational complexity while capturing both global and high-frequency features of the speech signal. To address the inherent insensitivity of T-MSA to channel information \cite{mb}, we introduce a streamlined channel and spatial attention branch called CSA. The features from these three branches are integrated to form the MET attention, which serves as the core module of the backbone network. Ultimately, amplitude and phase spectra are independently decoded, and the enhanced speech signal is reconstructed through Inverse Short-Time Fourier Transform (ISTFT).

Our model demonstrates remarkable effectiveness. Ultimately, through empirical validation on the VoiceBank-Demand dataset, MUSE achieved outstanding performance with extremely low parameter count of 0.51M and remarkably high utilization of memory(A 8GB GPU is sufficient for training).

\section{Proposed method}

In this section, we provide a comprehensive exposition of MUSE, which is founded upon a U-Net architectural framework, comprising dense convolution codec, deformable embedding, and MET-Transformer.

\subsection{Model architecture}

\begin{figure*}[t]
  \centering
  \includegraphics[width=\linewidth]{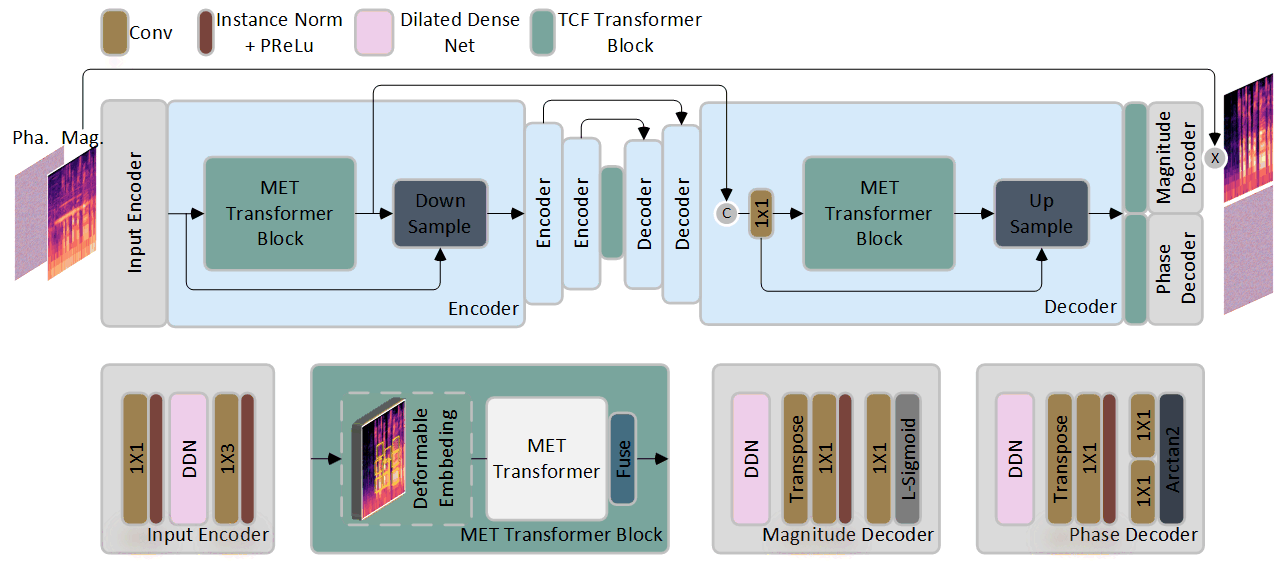}
  \caption{The overall architecture of MUSE.}
  \label{fig:speech_production}
\end{figure*}

we subject the input signal y to Short-Time Fourier Transform (STFT), yielding the magnitude spectrum
$\boldsymbol{Y}_{\boldsymbol{m}} \in \mathbb{R}^{T \times F}$
and the phase spectrum
$\boldsymbol{Y}_{\boldsymbol{p}} \in \mathbb{R}^{T \times F}$
.Following the methodology of PHASEN \cite{phasen}, a power-law compression is applied to
$\boldsymbol{Y}_{\boldsymbol{m}}$
, resulting in the compressed magnitude spectrum
${\boldsymbol{Y}_{\boldsymbol{m}}}^{c} \in \mathbb{R}^{T \times F}$
. The compressed ${\boldsymbol{Y}_{\boldsymbol{m}}}^{c}$ is concatenated with 
$\boldsymbol{Y}_{\boldsymbol{p}}$
to form
${\boldsymbol{Y}_{\boldsymbol{in}}}^{c} \in \mathbb{R}^{T \times F \times 2}$
, which is then fed into a U-Net architecture incorporating the MET-Transformer. For the ultimate decoding layer, independent feature extraction is performed on the magnitude spectra 
$\boldsymbol{\hat{X}}_{\boldsymbol{m}} \in \mathbb{R}^{T \times F}$
and phase spectra
$\boldsymbol{\hat{X}}_{\boldsymbol{p}} \in \mathbb{R}^{T \times F}$
. These individual features are subsequently fused into a spectrogram, and the signal is reconstructed through Inverse Short-Time Fourier Transform (ISTFT).

Following the input encoder layer, the MET encoder-decoder is introduced, with subsequent layers incorporating both upsampling and downsampling operations. Each stage comprises deformable embedding (DE) and MET-Transformer blocks. The DE yields tokens of varying scales for input into the Transformer. Channel expansion ${1d, 2d, 3d}$
is achieved for both encoders and decoders through downsampling and upsampling modules. In the ultimate layer, analogous to the structure of the initial encoder layer, independent learning is performed for the magnitude and phase spectra.

\subsection{Deformable Embedding}
In the realm of acoustic signal processing, sound frequencies and features exhibit significant variability. Voiceprints typically possess unique shapes, thus, convolutional receptive fields may capture specific voiceprint features inadequately. In pursuit of a more adaptable receptive field configuration, akin to strategies employed in the field of image processing, we introduce depthwise separable and deformable convolutions (DSDCN) \cite{mb}. Through the application of depthwise convolution and pointwise convolution, we mitigated computational complexity and reduced parameter count, consequently enhancing the efficiency of speech signal processing. 
We have incorporated Deformable Embedding at the head of each U-Net encoder-decoder, as illustrated by the MET Transformer block in Figure \ref{fig:speech_production}. DSDCN concurrently possesses the capability to learn intricate details of voiceprint information at fine scales and large-scale voiceprint features. This implies that DSDCN not only has the capability to capture the shape of voiceprint features but also exhibits the ability to expand the receptive field, similar to the behavior of dilated convolutions. Hardswish activation is employed to harness heightened non-linear characteristics, enabling the network to extract more intricate acoustic features.The application of DSDCN is avoided in the input encoder and magnitude-phase decoder of the network. This strategic exclusion arises from the inherent flexibility of deformable convolutions in achieving adaptable receptive fields through feature offsets, as the use of DCN across the full spectrum of the input and output codec stages is not efficient.

\subsection{MET-Transformer block}
\begin{figure}[t]
  \centering
  \includegraphics[width=\linewidth]{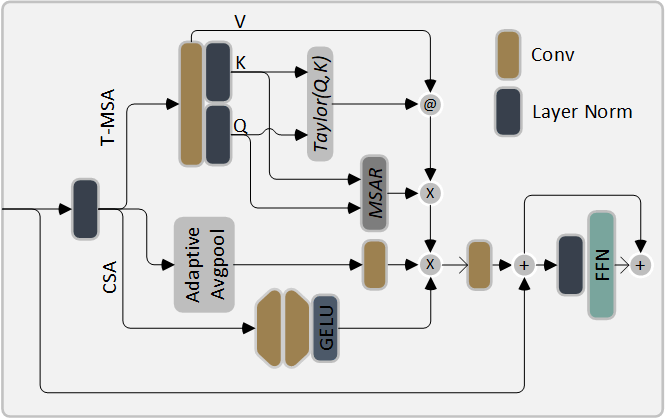}
  \caption{MET Transformer.}
  \label{fig:speech_production}
\end{figure}
In order to comprehensively learn the entire spectral characteristics of the speech signal, we prioritized efficiency in the design of the new Multi-path Enhanced Taylor (MET) Transformer block. MET is composed of three paths:
The first path incorporates Taylor multi-head attention (T-MSA). In comparison to MSA, T-MSA attenuates attention across channels, emphasizing global self-attention, and excelling in handling high-frequency information.
The second path, comprised of pooling and convolution linearly, is designed to compensate for channel information exchange in T-MSA.
The third path, a combination of pointwise and depthwise convolutions, is responsible for learning spatially-invariant features and maintaining the stability of information flow. In order to streamline the description, we designate the second and third paths as the  Channel and Spacial Attention (CSA) branch. Inspired by the 'Simple Gate' in NAFnet \cite{NAFnetchen2022simple}, we perform element-wise multiplication of these three branches, imparting a certain degree of non-linearity to our module.
In a manner akin to the structure of the Transformer, we sum the MET with the residual connection \cite{resnet}, subsequently linearly connecting it to the FFN module after layer normalization.

\textbf{T-MSA Branch.}
For vanilla Transformer, we have the following formula:
\begin{equation}\label{eq:1}
V^{\prime}=\operatorname{Softmax}\left(\frac{Q^T K}{\sqrt{D}}\right) V^T
\end{equation}
As the computation of softmax involves the natural logarithm of the exponentiated values $e^x$:
\begin{equation}\label{eq:2}
\operatorname{Softmax}\left(x_i\right)=\frac{e^{x_i}}{\sum_{i=1}^n e^{x_i}} \in(0,1)
\end{equation}
To approximate \eqref{eq:1} We can write a generalized function as:
\begin{equation}\label{eq:3}
V_i^{\prime}=\frac{\sum_{j=1}^N f\left(Q_i, K_j\right) V_j}{\sum_{j=1}^N f\left(Q_i, K_j\right)}
\end{equation}
where the matrix with '
$i$
' as the subscript represents the vector of the 
$i$
th row of a given matrix, and '
$f(\cdot)$
' denotes any similarity function, Equation \eqref{eq:3} reduces to Equation \eqref{eq:1} by setting $f\left(Q_i, K_j\right)=\exp \left(\frac{Q_i^T K_j}{\sqrt{D}}\right)$. To elucidate this transition, the application of the Taylor formula \cite{taylor1717methodus} for a first-order expansion at 0 allows us to write 
$Taylor\left(Q_i, K_j\right) = 1 + \frac{Q_i^T K_j}{\sqrt{D}} + o\left(\frac{Q_i^T K_j}{\sqrt{D}}\right) \approx \exp\left(\frac{Q_i^T K_j}{\sqrt{D}}\right)$and reformulate Equation \eqref{eq:3} as follows:
\begin{equation}\label{eq:4}
V_i^{\prime}=\frac{\sum_{j=1}^N\left(1+Q_i^T K_j+o\left(Q_i^T K_j\right)\right) V_j^T}{\sum_{j=1}^N\left(1+Q_i^T K_j+o\left(Q_i^T K_j\right)\right)}
\end{equation}

The computational complexity pertaining to both the MSA and the T-MSA, with respect to an spectrum comprising 
$t \times f$ 
patches, is explicated as follows:
\begin{equation}\label{eq:5}
\Omega(\mathrm{MSA})=4 t f D^2+2 t^2 f^2 D
\end{equation}
\begin{equation}\label{eq:6}
\Omega(\mathrm{T-MSA})=18 t f D+2 t f D^2
\end{equation}
Generally speaking, $D$ is significantly smaller than $tf$. From this, we can conclude that T-MSA can significantly reduce computational complexity while approaching MSA representation, especially as $tf$ increases, the reduction in complexity becomes more pronounced.

\textbf{CSA Branch.}
Differing from MSA, T-MSA accentuates global attention while attenuating channel attention \cite{mb}. We designed channel and spatial attention (CSA) branch. To guide the network towards inter-channel feature relationships and emphasize crucial channel information, we employ adaptive average pooling to calculate weights between channels \cite{wang2020eca}. In an effort to reduce network complexity and parameter count, we adopt a strategy inspired by NAFnet \cite{NAFnetchen2022simple}, retaining only the essential components of channel attention and directly employing 1x1 convolution operations for inter-channel information exchange. Notably, the element-wise multiplication operation of 3 branches introduces nonlinearity to MET like Simple Gate \cite{NAFnetchen2022simple}.
Within the spatial attention \cite{spatialattention} branch, we directly utilize pointwise and 3x3 depthwise convolutions \cite{dwconv} with GELU activation, incurring minimal computational cost and maintaining a low parameter count. This approach captures spatial information at different scales, aiding the network in focusing on crucial regions in both magnitude and phase spectra \cite{phasen}.

\subsection{Dense Convolution Codec}
In the input-output encoder, we borrowed the Codec from MP-SEnet \cite{mpsenet_interspeech}. The Dilated Dense-Net structure comprises four dilated convolutional blocks with dense residual connections \cite{dilateddensenet}, each block having dilation factors set to ${1, 2, 4, 8}$. In dilated convolutions, our primary objective is to augment the receptive field while preserving the kernel size and layer counts, ensuring the efficient capture of globally and intricately featured information.

\section{EXPERIMENTS}
\begin{table*}[th]
    \centering
    \caption{Comparison with other methods on VoiceBank+DEMAND dataset. ``-'' denotes the result is not provided in the original paper.}
    \label{tab:comparison}
    \setlength{\tabcolsep}{8pt} % 设置列之间的间隙为4pt
    \begin{tabular}{lccccccc}
        \toprule
        Method & Architecture & Parameters & PESQ & CSIG & CBAK & COVL & STOI \\
        \midrule
        Noisy & - & - & 1.97 & 3.35 & 2.44 & 2.63 & 0.91 \\
        SEGAN \cite{pascual2017segan}  & U-Net & 43.18M & 2.16 & 3.48 & 2.94 & 2.80 & 0.92 \\
        DEMUCS \cite{defossez2019demucs}  & U-Net & 33.53M & 3.07 & 4.31 & 3.40 & 3.63 & 0.95 \\
        SE-Conformer \cite{seconformer}  & U-Net & - & 3.13 & 4.45 & 3.55 & 3.82 & 0.95 \\
        MetricGAN+ \cite{metricgan+}  & LSTM & - & 3.15 & 4.14 & 3.16 & 3.64 & - \\
        TSTNN \cite{wang2021tstnn}  & Two-Stage Transformer & 0.92M & 2.96 & 4.33 & 3.53 & 3.67 & 0.95 \\
        DB-AIAT \cite{DB-AIATinterspeech12}  & Two-Stage Transformer & 2.81M & 3.31 & 4.61 & 3.75 & 3.96 & - \\
        DPT-FSNet \cite{DPT-FSNetinterspeech29}  & Two-Stage Transformer & 0.88M & 3.33 & 4.58 & 3.72 & 4.00 & \textbf{0.96} \\   
        PHASEN \cite{phasen} & Two-Stream DNN & 20.9M & 2.99 & 4.18 & 3.45 & 3.50 & 0.95 \\
        MetricGAN-OKDv2 \cite{metricganOKD} & LSTM &  0.82M & 3.12 & 4.27 & 3.16 & 3.71 & 0.95 \\
        MANNER \cite{park2022manner} & U-Net &  24.07M & 3.21 & 4.53 & 3.65 & 3.91 & 0.95 \\
        MANNER-S-5.3GF \cite{mannerlit} & U-Net &  0.90M & 3.06 & 4.42 & 3.58 & 3.77 & 0.95 \\
        \textbf{MUSE(Ours)} & U-Net & \textbf{0.51M} & \textbf{3.37} & \textbf{4.63} & \textbf{3.80} & \textbf{4.10} & 0.95 \\
        \bottomrule
    \end{tabular}
\end{table*}

\subsection{Datasets}

In our study, we employed the VoiceBank+DEMAND \cite{voicebank} dataset, which provides a comprehensive collection of high-fidelity utterances, both clean and mixed. The training set consists of 11,572 utterances, totaling 9.4 hours, delivered by 28 distinct speakers. Conversely, the test set comprises 824 utterances, amounting to 0.6 hours, articulated by 2 speakers not represented in the training data. Notably, the noise profiles in the test set, derived from recorded DEMAND datasets and synthetic sources, diverge from those present in the training set. Signal-to-noise ratios vary between the datasets, with the test set featuring ratios of 0dB, 5dB, 10dB, and 15dB, while the training set encompasses ratios of 2.5dB, 7.5dB, 12.5dB, and 17.5dB.
\subsection{Model setup}
Throughout the training process, speech data was uniformly segmented into 30700 points, employing an FFT size of 510, a window length of 510, a hop length of 100, and a sampling rate of 16000\footnote{{https://github.com/huaidanquede/MUSE-Speech-Enhancement}}. The training configuration encompassed a batch size of 2, a learning rate set to 0.0005 with a decay factor of 0.99, dense channels is 16, all the models were trained using the AdamW \cite{kingma2014adam} optimizer until 100 epochs. During training, a single 8GB RTX 3070ti GPU was utilized.
\subsection{Evaluation metrics}
We have selected a suite of widely accepted metrics to assess the quality of denoised speech. These metrics include the Perceptual Evaluation of Speech Quality (PESQ) \cite{pesq}, which provides scores ranging from -0.5 to 4.5, the Segmental Signal-to-Noise Ratio (SSNR), and Composite Mean Opinion Score (MOS) metrics as outlined in literature. The MOS metrics encompass the MOS prediction of signal distortion (CSIG), MOS prediction of background noise intrusiveness (CBAK), and MOS prediction of overall effect (COVL), each scored within a range of 1 to 5. Furthermore, Speech Transmission Index (STOI) \cite{stoi} is employed, offering scores from 0 to 1 to evaluate speech intelligibility, where higher values indicate superior performance across all metrics under consideration.
\subsection{Result}
\textbf{Comparative analysis} of objective metrics on the VoiceBank+DEMAND dataset, as presented in Table \ref{tab:comparison}. Due to MUSE is based on the U-Net, we selected 5 U-Net based method including SEGAN, DEMUCS, SE-Conformer, MANNER, and MANNER-S-5.3GF. Furthermore, we specifically focused on selecting lightweight models with a parameter count less than 1M for comparison, including TSTNN, DPT-FSnet, MetricGAN-OKDv2, and MANNER-S-5.3GF. Even with the inherent high parameter count of the U-Net architecture, MUSE achieved a parameter count of 0.51M, which is lower than all baseline models. Additionally, apart from slightly lower performance in STOI compared to DPT-FSNet, MUSE outperforms baseline models in terms of PESQ, CSIG, CBAK, and COVL metrics.
\begin{table}[htbp]
    \centering
    \caption{\textbf{Ablation Study:} CSA means Channel and Spatial Attention; T-MSA means Taylor-Multi-head Self Attention; DE means Deformable Embedding. We retrained the MP-SEnet with 16 dense channels as our baseline namely MP-SEnet-16.}
    \label{tab:ablation}
    \setlength{\tabcolsep}{5pt} % 设置列之间的间隙为4pt
    \begin{tabular}{lcccc}
    \toprule
    Model & PESQ & CSIG & CBAK & COVL \\
    \midrule
    MUSE   & \textbf{3.37} & \textbf{4.63} & \textbf{3.80} & \textbf{4.10} \\
    w/o CSA & 3.29 & 4.62 & 3.76 & 4.04 \\
    w/o T-MSA & 3.27 & 4.62 & 3.76 & 4.02 \\
    w/o DE & 3.34 & \textbf{4.63} & 3.78 & 4.08 \\
    MP-SEnet-16 (baseline) & 3.21 & 4.54 & 3.72 & 3.95 \\
    \bottomrule
    \end{tabular}%
\end{table}

\textbf{Ablation study} compared the scores of multiple objective metrics on the VoiceBank+DEMAND dataset, as shown in Table \ref{tab:ablation}. "w/o" denotes the removal of a certain component. We experimented on the major breakthroughs of MUSE, including the CSA branch, T-MSA branch, and Deformable Embedding in our proposed MET Transformer. MP-SEnet-16 represents a reproduced result of MP-SEnet \cite{mpsenet_interspeech} with only 16 dense channels, which serves as our original baseline model.
\section{Conclusions}
In this paper, we proposed a lightweight speech enhancement network named MUSE, achieved competitive performance with only 0.51M parameters. We firstly apply Deformable Embedding to flexibly adapt the shape of voiceprint. We propose a novel multi path enhanced Taylor-Transformer (MET) Transformer, utilizing pooling and convolutional branches to achieve channel information exchange and spatial attention missing in light weighted Taylor-Transformer. MUSE reduces training and deployment costs greatly. In the future, we will explore how to make MUSE into a real-time speech enhancement system, enabling it to have a broader range of applications.

\bibliographystyle{IEEEtran}
\bibliography{mybib}

\end{document}